\documentclass[sigplan,screen]{acmart}

\AtBeginDocument{%
  }

\usepackage{graphicx}
\usepackage{subcaption}
\usepackage{hyperref}
\usepackage{tikz}
\usepackage{xcolor}
\usepackage{amsmath}
\usepackage{listings}
\usepackage[frozencache,cachedir=.]{minted}

\def\toolname{daisy }

\setcopyright{cc}
\setcctype{by}
\acmDOI{10.1145/3696443.3708951}
\acmYear{2025}
\copyrightyear{2025}
\acmISBN{979-8-4007-1275-3/25/03}
\acmConference[CGO '25]{Proceedings of the 23rd ACM/IEEE International Symposium on Code Generation and Optimization}{March 01--05, 2025}{Las Vegas, NV, USA}
\acmBooktitle{Proceedings of the 23rd ACM/IEEE International Symposium on Code Generation and Optimization (CGO '25), March 01--05, 2025, Las Vegas, NV, USA}
\received{2024-09-11}
\received[accepted]{2024-11-04}

\begin{document}

\title{A Priori Loop Nest Normalization: Automatic Loop Scheduling in Complex Applications}

\author{Lukas Trümper}
\orcid{0000-0002-0961-7723}
\affiliation{%
  \institution{Daisytuner}
  \city{Darmstadt}
  \country{Germany}
}
\email{lukas.truemper@daisytuner.com}

\author{Philipp Schaad}
\orcid{0000-0002-8429-7803}
\affiliation{%
  \institution{ETH Zurich}
  \city{Zurich}
  \country{Switzerland}
}
\email{philipp.schaad@inf.ethz.ch}

\author{Berke Ates}
\orcid{0000-0003-0242-3640}
\affiliation{%
  \institution{ETH Zurich}
  \city{Zurich}
  \country{Switzerland}
}
\email{beates@student.ethz.ch}

\author{Alexandru Calotoiu}
\orcid{0000-0001-9095-9108}
\affiliation{%
  \institution{ETH Zurich}
  \city{Zurich}
  \country{Switzerland}
}
\email{alexandru.calotoiu@inf.ethz.ch}

\author{Marcin Copik}
\orcid{0000-0002-7606-5519}
\affiliation{%
  \institution{ETH Zurich}
  \city{Zurich}
  \country{Switzerland}
}
\email{marcin.copik@inf.ethz.ch}

\author{Torsten Hoefler}
\orcid{0000-0002-1333-9797}
\affiliation{%
  \institution{ETH Zurich}
  \city{Zurich}
  \country{Switzerland}
}
\email{htor@inf.ethz.ch}

\begin{abstract}
The same computations are often expressed differently across software projects and programming 
languages.
In particular, how computations involving loops are expressed varies due to the many possibilities to permute and compose loops.
Since each variant may have unique performance properties, automatic approaches to loop scheduling must support many different optimization recipes.
In this paper, we propose a priori loop nest normalization to align loop nests and reduce the variation before the optimization.
Specifically, we define and apply normalization criteria, mapping loop nests with different memory access patterns to the same canonical form.
Since the memory access pattern is susceptible to loop variations and critical for performance, this normalization allows many loop nests to be optimized by the same optimization recipe.
To evaluate our approach, we apply the normalization with optimizations designed for only the canonical form, improving the performance of many different loop nest variants.
Across multiple implementations of 15 benchmarks using different languages, we outperform a baseline compiler in C on \textit{average} by a factor of $21.13$, state-of-the-art auto-schedulers such as \textit{Polly} and the \textit{Tiramisu auto-scheduler} by $2.31$ and $2.89$, as well as performance-oriented Python-based frameworks such as \textit{NumPy}, \textit{Numba}, and \textit{DaCe} by $9.04$, $3.92$, and $1.47$.
Furthermore, we apply the concept to the \textit{CLOUDSC} cloud microphysics scheme, an actively used component of the Integrated Forecasting System, achieving a 10\% speedup over the highly-tuned Fortran code.
\end{abstract}




\begin{teaserfigure}
    \centering
  \includegraphics[width=0.9\linewidth]{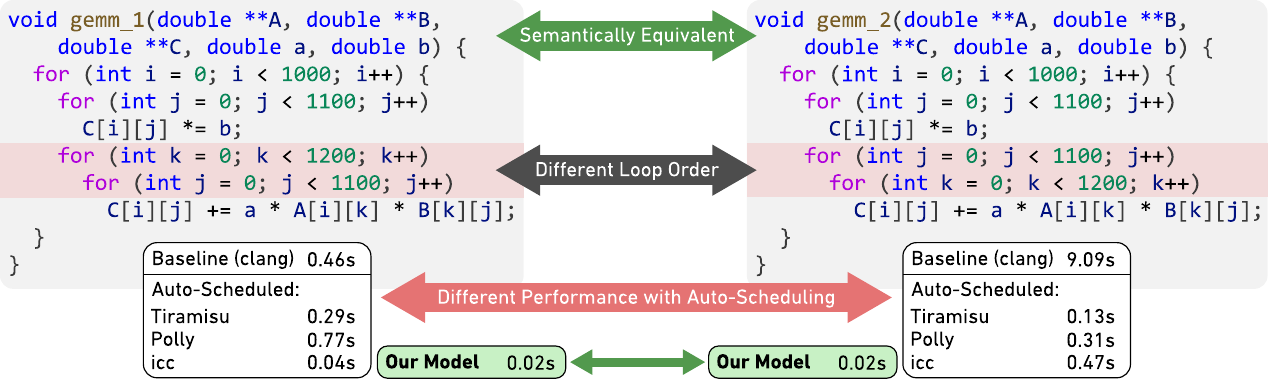}
  \caption{Structurally different General Matrix-Matrix Multiply (GEMM) kernels yield significantly different performance.}
  \Description{.}
  \label{fig:posterchild}
\end{teaserfigure}


\maketitle

%\newpage

\section{Introduction}
\label{sec:Introduction}

Because of sophisticated code optimizations by compilers, developers of high-performance software applications can often conveniently express the same linear code differently while achieving high performance on modern processors.
However, the performance of loop-based computations is highly susceptible to the specific permutations and composition of loops chosen by the developer.

Different frameworks implement automatic loop scheduling methods to enable automatic high-performance loop-based code.
Since loop scheduling is a complex problem in and of itself~\cite{Adams:2019,Steiner:2021}, these frameworks rely on approximate methods.
For instance, \textit{Polly}~\cite{Grosser:2011} finds a good schedule by optimization of an integer-linear program (ILP)~\cite{Bondhugula:2008b}.
However, \citet{Baghdadi:2019} show that this ILP only covers a certain fraction of the practically-relevant scheduling space~\cite{Baghdadi:2019}.
Recent approaches~\cite{Adams:2019,Baghdadi:2021} leverage deep learning to search much larger scheduling spaces at the price of local optima.

Although such auto-schedulers can achieve significant speedups, their direct application to large, scientific applications is currently limited.
The small example of GEMM presented in Figure~\ref{fig:posterchild} already shows that the results of auto-schedulers may vary by factors of $3\times$-$10\times$ depending on the chosen loop order.
Hence, developers must manually align loop nests to look like the supported optimization recipes.

To enable robust, automatic loop scheduling in complex applications, this paper introduces a priori loop nest normalization for auto-scheduling.
Specifically, we identify with \textit{maximal loop fission} and \textit{stride minimization} two normalization criteria, which canonicalize loop nests with different memory access patterns.
This method allows us to more easily apply the same optimization recipes to various loop nests with different performance properties.
We test the robustness of the optimization plus normalization with multiple semantically equivalent implementations of 15 benchmarks from \textit{PolyBench}~\cite{Pouchet:2017} across Python and C and with the highly optimized cloud microphysics scheme CLOUDSC written in Fortran.
This results in our optimization pipeline outperforming not just baseline compiler results in C ($\times21$) and Fortran ($\times1.1$), but also non-normalizing state-of-the-art auto-schedulers such as Polly~\cite{Grosser:2011} ($\times2.3$), and the \textit{Tiramisu auto-scheduler}~\cite{Baghdadi:2021} ($\times2.9$), as well as performance-oriented Python-based frameworks such as \textit{NumPy}, \textit{Numba}, and \textit{DaCe} by $9.04$, $3.92$, and $1.47$.
In short, our contributions are
\begin{itemize}
    \item Definition of a priori loop nest normalization to align loop nests with different performance properties.
    \item Implementation for LLVM IR and integration with a state-of-the-art loop scheduling algorithm.
    \item Evaluation of the robustness of optimization plus normalization on representative benchmarks across multiple programming languages as well as a case study optimizing a highly tuned cloud micro-physics simulation.
\end{itemize}

\section{A-Priori Loop Nest Normalization for Auto-Scheduling}
\label{sec:Normalization}

The cost of moving data through the memory hierarchy is the dominant factor for the performance of modern computing architectures~\cite{Unat:2017}.
Therefore, a priori loop nest normalization aims to provide simple memory access patterns as the starting point of the optimization.
It should map loop nests with vastly different performance properties such as the \textit{reuse distance}~\cite{Coffman:1973, Beyls:2001, Schaad:2022} or the \textit{sustained memory bandwidth} to the same canonical form.
After formally defining what we understand as loops and computations, we will introduce the two normalization criteria based on well-known compiler transformations.

\begin{figure}[ht]
    \begin{subfigure}{0.4\textwidth}
    \centering
    \includegraphics[width=0.45\columnwidth]{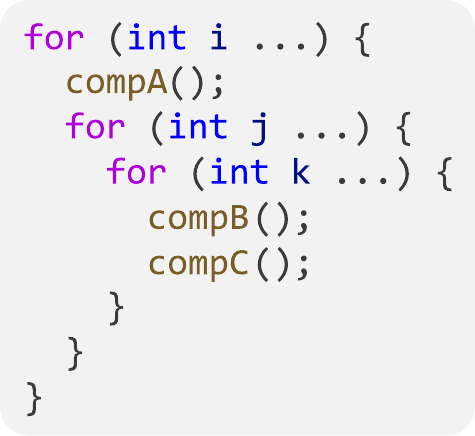}
    \caption{Loop nest pseudocode}
    \label{fig:Loop_Nest_pseudocode}
    \vspace{0.25cm}
    \end{subfigure}
    \begin{subfigure}{0.4\textwidth}
    \centering
    \includegraphics[width=0.45\columnwidth]{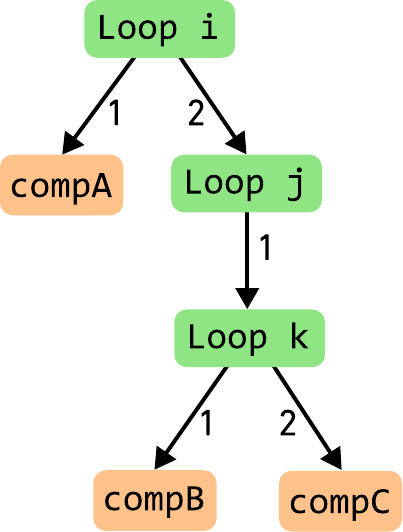}
    \caption{Loop nest tree representation}
    \end{subfigure}
    \caption{Characterization of loop nests.}
    \label{fig:loop_nest}
\end{figure}

\paragraph{Computation.}
We define a \textit{computation} as a unit of work composed of one or more instructions, where exactly one of the instructions is a write of a scalar value to a data container.

\paragraph{Loop.}
A \textit{loop} comprises an iterator with its initial values and update criterion, a termination condition, and a loop body composed of a sequence of computations. 

\begin{figure*}[ht]
    \centering
    \begin{subfigure}[b]{.4\linewidth}
        \centering
        \includegraphics[width=0.8\columnwidth]{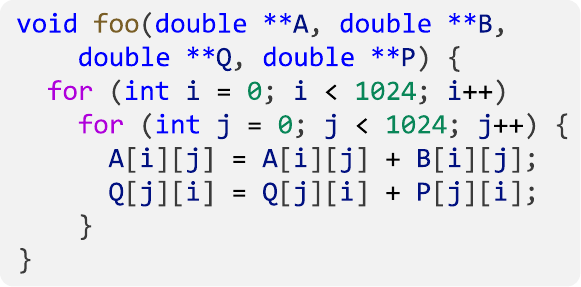}
        \vspace{.5em}
        \caption{Two independent computations with contiguous\\and strided memory accesses in a single loop.}
        \label{fig:foo_init}
    \end{subfigure}
    \begin{subfigure}[b]{.4\linewidth}
        \centering
        \includegraphics[width=0.8\columnwidth]{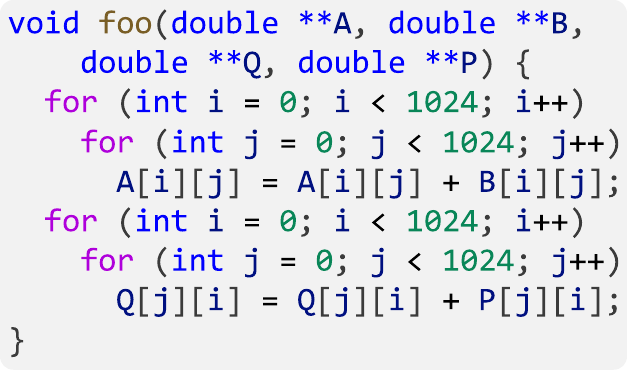}
        \caption{The same computations fissioned into two separate loop nests.}
        \label{fig:foo_fission}
    \end{subfigure}
    \vfill
    \begin{subfigure}[b]{0.8\linewidth}
        \centering
        \vspace{0.25cm}
        \includegraphics[width=\linewidth]{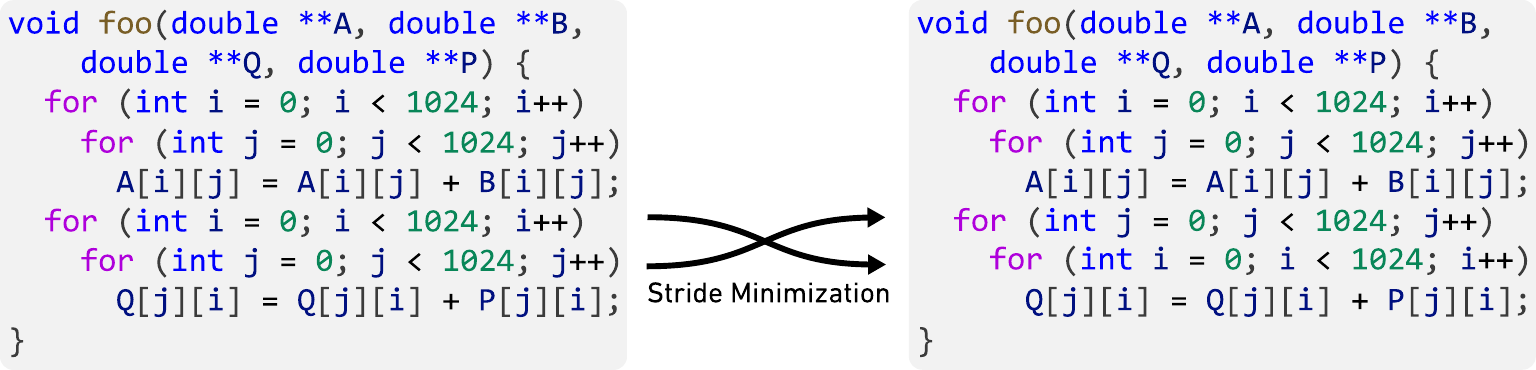}
        \caption{Permutation of the second loop nests to minimize the strides of memory accesses.}
        \label{fig:foo_stride}
        \vspace{-0.25cm}
    \end{subfigure}
    \caption{Loop nest code samples subject to normalization.}
    \label{fig:fission_strides}
\end{figure*}

\paragraph{Loop nest.}
A \textit{loop nest} is a loop where the loop body can be composed by an ordered sequence of computations, loops, and loop nests.
In the following, we use the notation $comp[i, j, k]$ if a computation is nested inside the loops $i$, $j$, and $k$, where $i$ is the outermost and $k$ the innermost loop.
Similarly, we use the notation $loop[i, j]$ to represent a loop nest or loop nested within the loops $i$ and $j$.
Figure~\ref{fig:loop_nest} illustrates the tree representation of a loop nest.

\subsection{Maximal Loop Fission}
\label{subsec:Normalization_Fission}

Fusing computations into shared loops is a common technique to improve performance~\cite{Mehta:2014}.
However, the combination of computations usually increases the complexity of memory accesses.
% For instance, when fusing hierarchical stencils, each additional computation increases the offset to be read in different dimensions of an array.
An example, shown in Figure~\ref{fig:foo_init}, combines two computations with contiguous and strided memory accesses.
Since the developer may apply such compositions manually when writing the code, we propose simplifying all loops to make them easier to analyze:
we fission them as much as possible before the optimization

Let $comp1[i_1, ..., i_j, ..., i_n]$ and $comp2[i_1, ..., i_j, ..., i_n]$ be two computations within the same loop nest.
If there are no data dependencies or loop-carried dependencies between $comp1$ and $comp2$, we define a new loop nest with the same iterator, initial values, update criteria, and termination conditions $i_1'=i_1, ..., i_n'=i_n$.
We then divide $comp1$ and $comp2$ across loop nests such that after fissioning we have $comp1[i_1, ..., i_j, ..., i_n]$ and $comp2[i_1', ..., i_j', ..., i_n']$.
We repeat this for loops and computations nested at \textit{the same level} of every loop nest until no such transformations are possible.
The result is a sequence of "atomic" loop nests, as their loop bodies contain computations and loops that can not be separated due to data dependencies.
An example of one instance of fissioning is illustrated in Figure~\ref{fig:foo_init} and Figure~\ref{fig:foo_fission}, where the two computations are split into separate loop nests.

\subsection{Stride Minimization}
\label{subsec:Normalization_Strides}

Depending on the order of loops within a loop nest, memory accesses have different strides, impacting the cache utilization.
Since the optimal order depends on other optimizations such as tiling or vectorization, loop permutation requires a non-trivial performance model and must be decided by an auto-scheduler~\cite{Mezdour:2023}.
To reduce the variations of loop nests, we propose stride minimization as a normalization criterion before optimization. 
We assume the stride minimization criterion is applied after the maximal loop fission criterion.

Let $loop \rightarrow (i_1, ..., i_j,i_k ..., i_n)$ be a loop nest with all iterators of nested loops ordered according to an in-order traversal.
We define a generic optimization criterion, $stride(loop)$, which maps subsequent accesses to arrays within each computation of a loop nest $loop$ to a real value.
For instance, the sum of all distances between two subsequent accesses to all arrays over all computations is a suitable function.

Let $\pi_1(loop) = (i_1, ..., i_j,i_k ..., i_n)$ be a permutation and $\pi_2(loop) = (i_1, ...,i_k, i_j, ..., i_n)$ be another legal permutation of the same loop nest.
If $stride(\pi_1)<stride(\pi_2)$, then $\pi_1$ is the permutation with the smaller stride.
Generalizing, for each loop nest $loop_{min}$, we find and replace it with the legal permutation with a minimal stride $\pi_{min}$.

The complexity of finding $loop_{min}$ depends on the definition of $stride(loop)$.
Since our goal is to reduce the variation of loop nests for a downstream auto-scheduler, we argue that the minimum can simply be found by enumeration for many practically-relevant loop nests.
For deep loop nests, we propose to sort groups of iterators as an approximation.

An example of stride minimization can be seen in Figure~\ref{fig:foo_stride}, where the the sum of strides in minimized.
If the dimensions are not statically known, other definitions of $stride(loop)$ must be used, e.g., the number of out-of-order access w.r.t. the permutation of loop iterators and array dimensions.

\begin{figure*}[ht]
    \centering
    \includegraphics[width=0.85\linewidth]{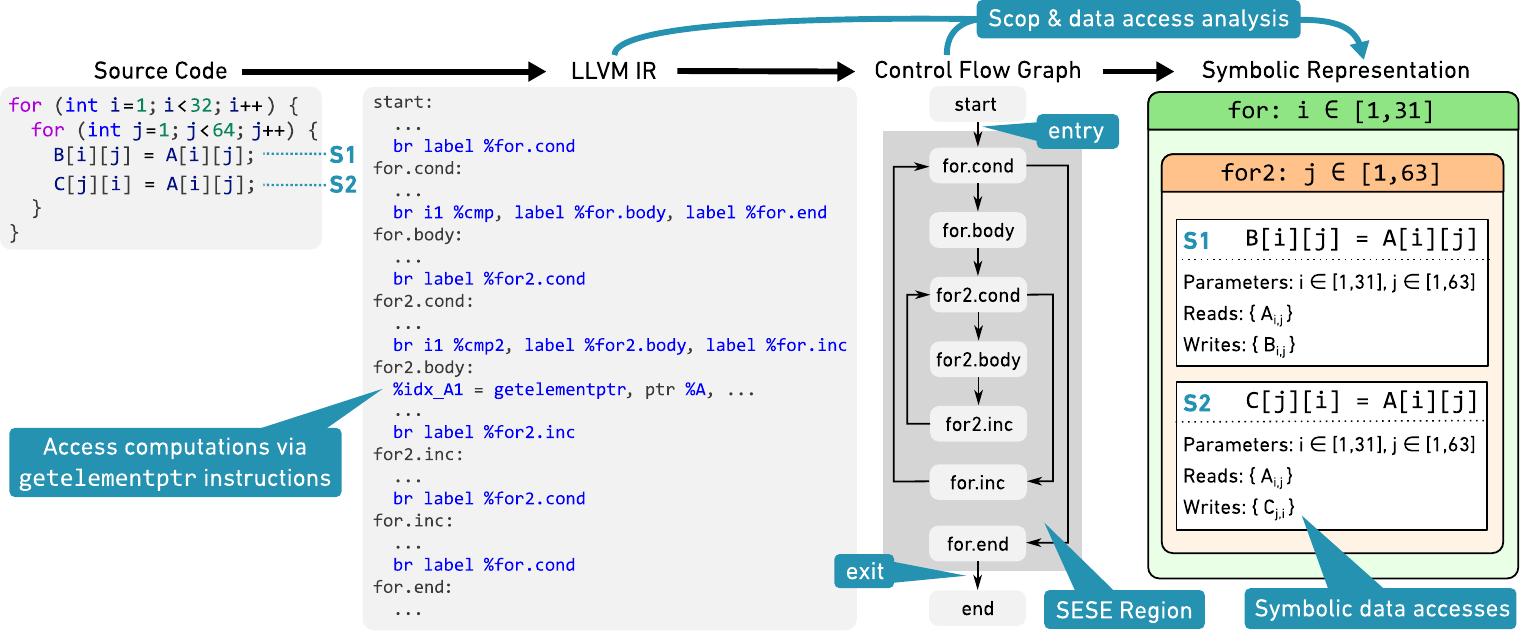}
    \caption{Lifting a symbolic representation of loop nests with high-level information from source code translated to LLVM IR.}
    \Description{.}
    \label{fig:lifting}
\end{figure*}

\section{Normalization on Intermediate Representations}
\label{sec:Implementation}

% Many auto-schedulers such the Tiramisu auto-scheduler~\cite{Baghdadi:2021} define their loop optimizations on a domain-specific language (DSL), providing all necessary high-level information.
We implement the normalization on LLVM IR to apply to as many codes as possible.
In LLVM IR, loops and memory accesses are represented as instructions.
Hence, all high-level information, such as array shapes, loop relations, and data dependencies, must be inferred through static analysis.
For instance, to identify the strided memory access to array \texttt{C} in computation \texttt{S2} of Figure~\ref{fig:lifting}, the accesses must first be derived from \texttt{store} and \texttt{getelementptr} and \texttt{branch} instructions.
Therefore, we first lift a rich, symbolic representation of loop nests from LLVM IR to do the normalization.
This lifting aims to represent loop nests in a hierarchy of loop and computation nodes, where loop iterators, domains, and data accesses are symbolic expressions.
Figure~\ref{fig:lifting} depicts the basic idea of this workflow.

\subsection{Lifting Symbolic Representations from LLVM IR}
\label{subsec:Lifting}

LLVM IR consists of instructions grouped into basic blocks.
Basic blocks are connected through conditional and unconditional branches, typically represented in a \textit{control-flow graph (CFG)}.
Polly implements several mechanisms to detect and lift a polyhedral representation of loop nests from LLVM IR.
Since this simplifies large parts of the symbolic analysis, we use Polly as the basis of our lifting workflow.
Based on Polly's representation, we then generate an \textit{Abstract Syntax Tree (AST)} of the loop nest using existing methods~\cite{Grosser:2015} implemented in the \textit{integer set library}~\cite{Verdoolaege:2010}.
This AST consists of a a tree of loops and computations nodes similar to our definition.
To analyze strides and fissioning opportunities efficiently, we further augment the tree with dataflow information describing the subset of data produced and consumed by different nodes.
We use the Stateful DataFlow multiGraph (SDFG)~\cite{BenNun:2019} and existing dataflow analysis~\cite{Calotoiu:2022}.

To apply the changes to the original code, we use the property that the loop nests detected by Polly are maximal \textit{single-entry-single-exit (SESE) regions}~\cite{Johnson:1994}.
SESE regions are subgraphs of a CFG with unique incoming and outgoing edges.
SESE regions can easily be removed from the original code and replaced by a function call to an external source code generated from an SDFG.

\subsection{Normalization Passes}
\label{subsec:Normalization_Passes}

\begin{figure*}[ht]
    \centering
    \includegraphics[width=0.9\linewidth]{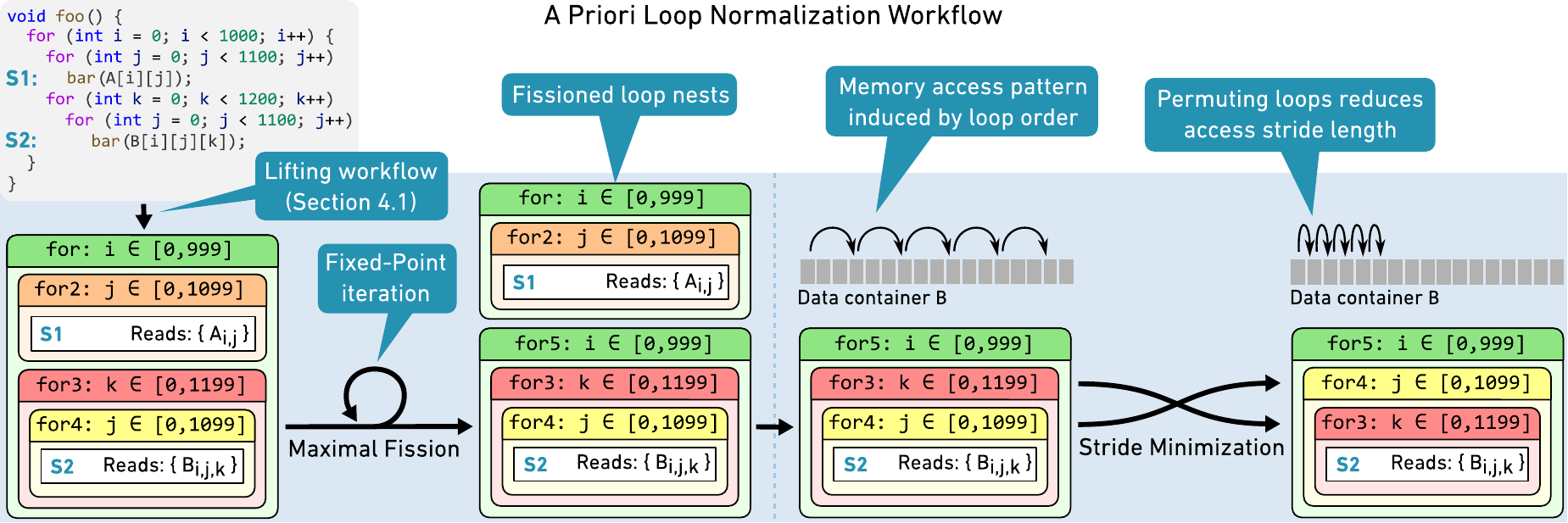}
    \caption{The normalization pipeline in two steps: Maximal loop fission and stride minimization.}
    \Description{.}
    \label{fig:normalize}
\end{figure*}

We implement the two normalization criteria as two separate transformation passes in a pipeline based on the lifted representation of loop nests.
An overview of the normalization pipeline is shown in Figure~\ref{fig:normalize}.
In the first step, we fission the loop nests as maximally as possible.
Since loop fissioning always splits loop nests into smaller loop nests of fewer computation nodes, we can apply transformations in a fixed-point pipeline until no more fissioning transformations apply.
In the second step, we search for the loop permutation with minimal strides for each resulting loop nest of the first step.
To find a minimal permutation, we enumerate all permutations and compute the strides from the symbolic expressions of memory accesses.
It should be noted that although our representation is lifted from LLVM IR, the normalization can also be applied to SDFGs obtained from other sources.

\section{The \toolname scheduler}
\label{sec:AB_Testing}

Ideally, normalization should make semantically equivalent implementations of an algorithm performance-equivalent.
To test this hypothesis, we create a normalized auto-scheduler, \textit{daisy}, and evaluate the impact of normalization on the auto-scheduler robustness.
We create an A/B test comparing the original implementation of benchmarks (A variants) and alternative, semantically equivalent implementations (B variants). 

\paragraph{Optimization Algorithm}
We define a new auto scheduler, which applies our normalization passes and then queries optimizations from a database using \textit{similarity-based transfer tuning}~\cite{Truemper:2023}.
The stride minimization uses the sum of strides of all array accesses as the optimization criterion.
The database consists of pairs of an embedding for the loop nest and transformation sequences including loop interchange, tiling, parallelization and vectorization.
The database is seeded from normalized loop nests of the A variants and then applied to the normalized B variants.
If a B loop nest is not reduced to an A loop nest, the transformation sequence cannot be applied.

\paragraph{Seeding a Scheduling Database}
We collect all loop nests from the normalized A variants to define the auto-scheduler.
For each loop nest corresponding to a BLAS-3 kernel, we add an optimization recipe to perform idiom detection, i.e., replacing the loop nest with the matching BLAS library call.
The optimizations for other loop nests are found using an evolutionary search.
In the first epoch of the search, the candidate optimizations for each loop nest are seeded using the Tiramisu auto-scheduler.
This population is refined in three iterations through standard mutation and selection techniques, where the runtime determines the fitness.
In the second and third epochs, the population is re-seeded using the current best optimization of the ten most similar loop nests and refined in three iterations again.
The Euclidean distance of \textit{performance embeddings}~\cite{Truemper:2023} determines the most similar loop nests.

\paragraph{Benchmarks}
PolyBench~\cite{Pouchet:2017} is a popular set of benchmarks for evaluating polyhedral compilers and auto-scheduling methods.
Many of the implemented benchmarks offer several degrees of freedom, where the loop nests can be nested and permuted differently without changing the semantics of the algorithm.
We have selected 15 parallelizable benchmarks where schedulers have a significant search space for optimization.
To evaluate the auto-scheduler robustness, we randomly generate an alternative B variant for each benchmark based on different permutations and compositions.
In the following, we only consider the large input size.

\paragraph{Baselines}
We use Polly~\cite{Grosser:2011} based on LLVM 16.0.4 and with optimization flags \textit{-O3 -polly -polly-parallel -polly-tiling -polly-vectorizer=stripmine -polly-2nd-level-tiling -gomp}.
We configure the Tiramisu auto-scheduler~\cite{Baghdadi:2021} to run a Monte-Carlo Tree Search guided by the performance model.
To account for the stochasticity of this search, we test the top three candidates and apply the best optimization among these.
Building the original Tiramisu auto-scheduler for the Tiramisu DSL~\cite{Baghdadi:2019} using currently available software packages was unsuccessful.
Therefore, we run the auto-scheduler as a standalone search and implement an adapter, which converts SDFGs to the JSON representation by the search.
To simplify the conversion, we apply the maximal loop fission criterion as part of the adapter and restrict the conversion to perfectly nested parallel loops.
% Although this may introduce a bias in terms of speedups, the robustness of the model to loop nest variations can still be analyzed since differently permuted A and B variants yield different JSON representations.
Moreover, we compare the results to icc 2021.9.0, a general baseline with auto-parallelization \textit{-parallel} and optimization \textit{-O3} flags enabled.

\paragraph{Experimental Setup}
The experiments are performed on an Intel Xeon E5-2680v3 clocked at 2.50 GHz with 64 GB of main memory.
We measure according to a standard framework~\cite{Hoefler:2015}, where measurements are taken until the variance drops below five percent, and the resulting median is reported as the runtime.

\subsection{Normalized Auto-Scheduling: Same Semantics, Same Performance}
\label{subsec:Normalized_Scheduling}

\begin{figure*}
    \includegraphics[width=0.9\linewidth]{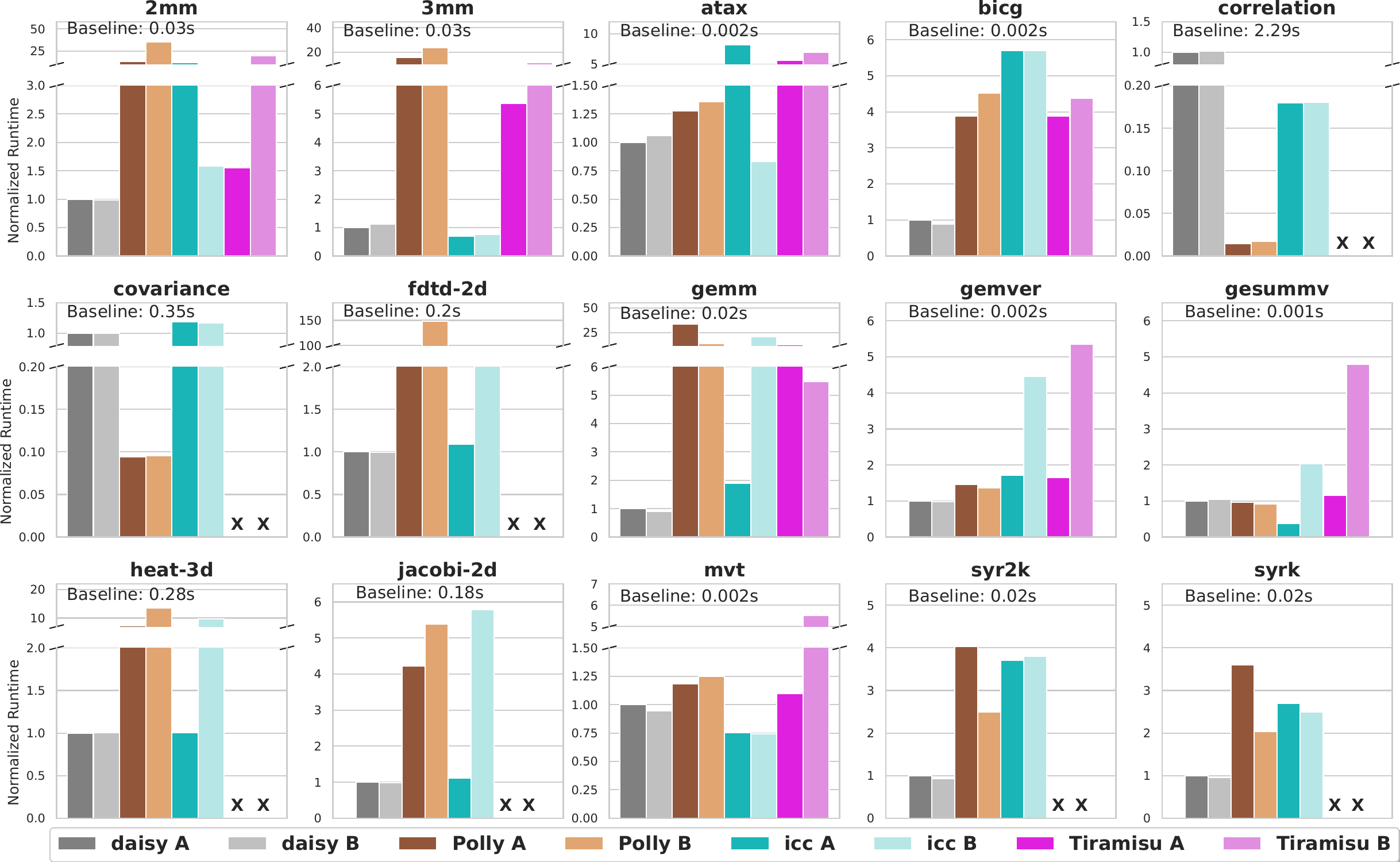}
    \caption{Comparison of our model with state-of-the-art auto-scheduling methods and the icc compiler. The runtime is expressed relative to the runtime of the A variant of the benchmarks using \textit{daisy}. Hence, a lower value is better. The implementation of the Tiramisu scheduler could not be applied to some of the benchmarks successfully. We mark those with \textbf{X}.}
    \label{fig:PolyBench_TT}
\end{figure*}

We evaluate the runtime of Polly, Tiramisu, and icc against \toolname for the two implementations - A and B - of each of the 15 benchmarks. The results are summarized in Figure~\ref{fig:PolyBench_TT}.

\paragraph{Robustness}
The first observation is that since the A and B variants are semantically equivalent, a robust auto-scheduler should achieve a runtime ratio close to one.
This is true \textbf{for \toolname, where the largest difference between the performance of the A and B implementations is 14\%} and the \textbf{mean difference is just 5\%}.
However, all other approaches show significant variation between A and B implementations on several benchmarks, with differences of over an order of magnitude for applications such as \textit{2mm} or \textit{fdtd-2d}.
For the latter, this can be explained by strided memory accesses in the B implementation that neither Polly nor icc can optimize well.
Similarly, the performance of the Tiramisu auto-scheduler is susceptible to the specific structure of the loop nests of the A and B variants.

\paragraph{Performance}
While achieving the same performance for the different implementations of the same benchmark is vital to prove the robustness of a scheduler, the performance must also be competitive with state-of-the-art approaches.
Figure~\ref{fig:PolyBench_TT} shows the runtime of each benchmark and the baseline auto-schedulers relative to \textit{daisy}.
Our model achieves a geometric mean speedup of $2.31$ over Polly, $2.89$ over the Tiramisu auto-scheduler, and $1.58$ over icc on the A variants.
For the B variants, our auto-scheduler achieves a geometric mean speedup of $2.97$ over Polly, $7.03$ over the Tiramisu auto-scheduler, and $2.51$ over icc.
Our model underperforms compared to Polly on \textit{correlation} and \textit{covariance}.
In those cases, our normalization passes fail to lift specific loop nests to the symbolic representations.
As a result, the loop nest is not optimized, and a reduction is executed in parallel, causing expensive atomic reductions in the C++ code.
In general, however, \toolname proves to be a sufficiently complex auto-scheduler.
Most importantly, the runtime variations across A and B variants are in the order of measurement noise for our model.
Hence, \toolname optimizes the A and B variants of the benchmarks equally well using only optimizations derived from normalized A variants.

\subsection{Ablation Study: Same Optimizations, Different Performance}

\begin{figure*}
    \includegraphics[width=0.9\linewidth]{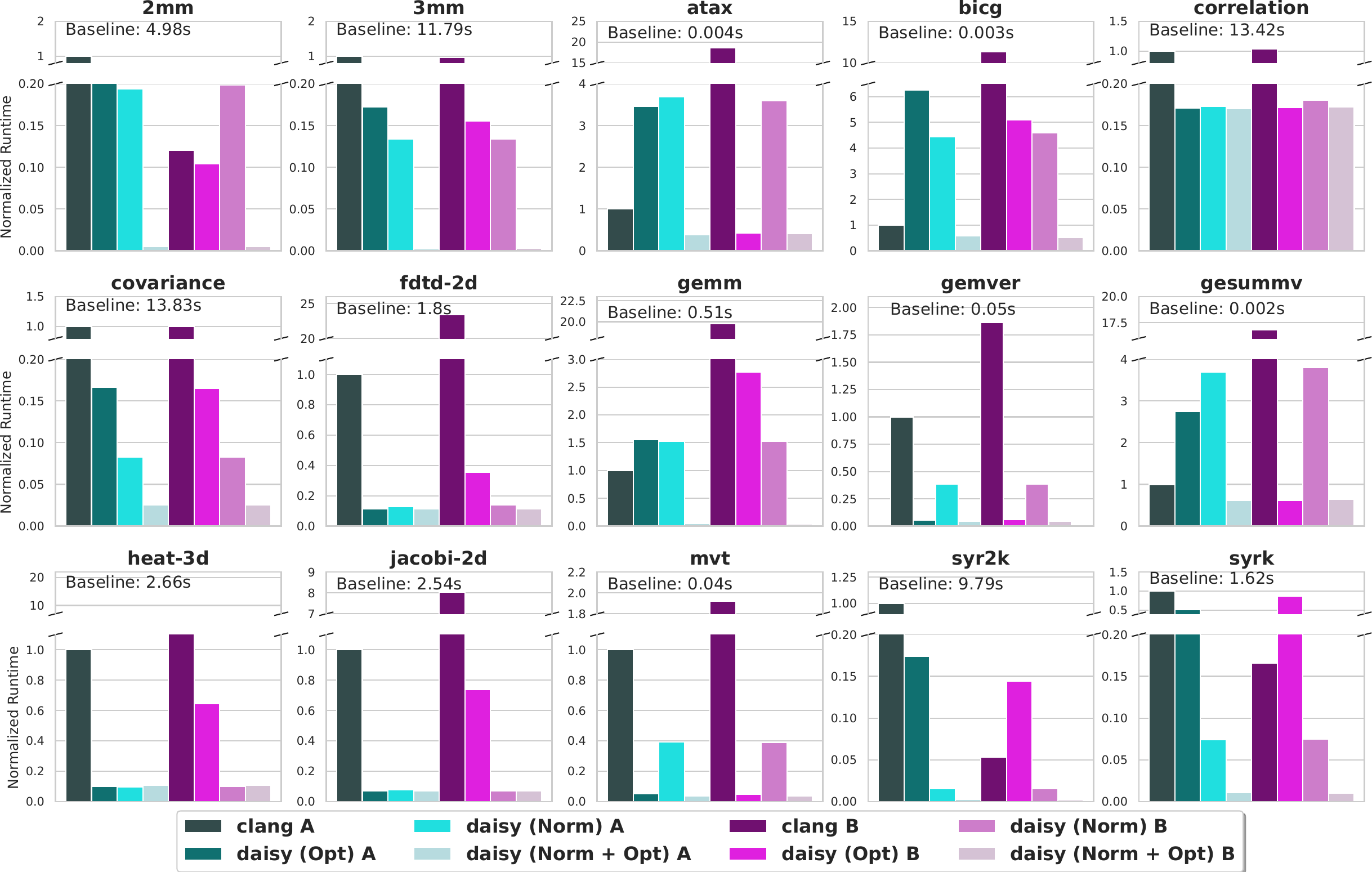}
    \caption{Comparison of clang and our model with and without normalization. The runtime is expressed relative to A variants of the benchmarks using clang. Hence, a lower value is better.}
    \label{fig:PolyBench_AB_No_normalization}
\end{figure*}

To analyze the impact of the normalization and the optimizations offered by transfer tuning in isolation, we now compare the results of compiling the benchmarks in the following scenarios: using only clang, using transfer tuning without normalization, using normalization without transfer tuning, and finally using the full pipeline in \toolname.
We do this for both the A and B versions of each benchmark. We note that the normal compiler optimizations \texttt{-O3}, etc. are applied in all configurations.
Figure~\ref{fig:PolyBench_AB_No_normalization} summarizes the relative runtime of the benchmarks for optimization with and without prior normalization. The results show that \textbf{both normalization and optimization using the similarity-based transfer tuning algorithm are required to reach the best performance consistently}.
Without the normalization step, the database queried by the transfer tuning algorithm would need to explicitly enumerate all possible loop variations, which would not scale.
% Thus, a priori normalization reduces the necessary model complexity while still achieving great performance, which is a typical indication of improved robustness according to \textit{Occam's razor}~\cite{Bargagli:2022}.

\subsection{Auto-Scheduling beyond C: Different Language, Same Optimization}

Applying auto-schedulers across programming languages is essential -- and we wish to test \toolname not just on different C implementations but also on implementations of the \textbf{same benchmarks in Python} -- increasing the number of implementation variants considered.
However, different programming languages have different syntactical features.

\begin{figure}[ht]
\centering
\begin{subfigure}{0.49\textwidth}
\centering
\includegraphics[width=0.85\columnwidth]{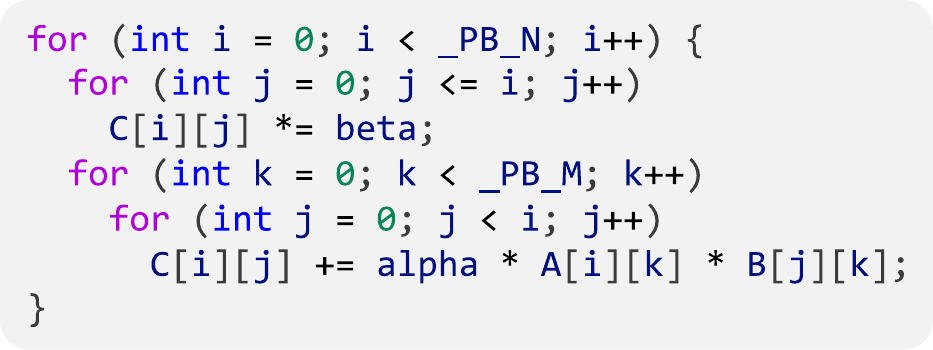}
% \begin{minted}[fontsize=\footnotesize]{c}
% for (i = 0; i < _PB_N; i++) {
%     for (j = 0; j <= i; j++)
%         C[i][j] *= beta;
%     for (k = 0; k < _PB_M; k++)
%         for (j = 0; j <= i; j++)
%             C[i][j] += alpha * A[i][k] * A[j][k];
% }
% \end{minted}
\caption{PolyBench's SYRK in C}
\end{subfigure}\hfill
\begin{subfigure}{0.49\textwidth}
\centering
\includegraphics[width=0.85\columnwidth]{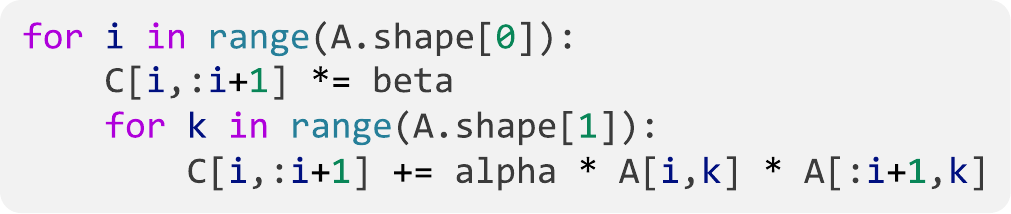}
% \begin{minted}[fontsize=\footnotesize]{python}
%   for i in range(A.shape[0]):
%       C[i, :i + 1] *= beta
%       for k in range(A.shape[1]):
%           C[i, :i + 1] += alpha * A[i, k] \
%                           * A[:i + 1, k]
% \end{minted}
\caption{NPBench's SYRK using NumPy}
\end{subfigure}
\caption{The SYRK kernel implemented in C and NumPy.}
\label{fig:SYRK}
\vspace{-1em}
\end{figure}

For instance, Figure~\ref{fig:SYRK} shows the implementations of the \textit{symmetric rank-k update (SYRK)} kernel in PolyBench~\cite{Pouchet:2017} and NPBench~\cite{Ziogas:2021}, a scientific benchmarking suite for high-performance NumPy~\cite{Harris:2020}.
For SYRK, the NPBench implementation uses ranges for indexing of NumPy arrays.
When translating the Python benchmarks to our IR, such syntactical features may yield different representations for the same programs because of additional translation and optimization passes.
To evaluate the effect of a priori loop nest normalization for auto-scheduling across programming languages, we apply the same database-based auto-scheduler from Section \ref{subsec:Normalized_Scheduling} to PolyBench benchmarks implemented in NPBench.
In detail, we use the DaCe Python frontend~\cite{BenNun:2019} to obtain an SDFG for the NPBench benchmark and then apply the normalization and auto-scheduler analogously.
For reasons of comparability, we adapt the input sizes of the NPBench benchmarks to the large variants of PolyBench.
\begin{figure*}
    \includegraphics[width=0.9\linewidth]{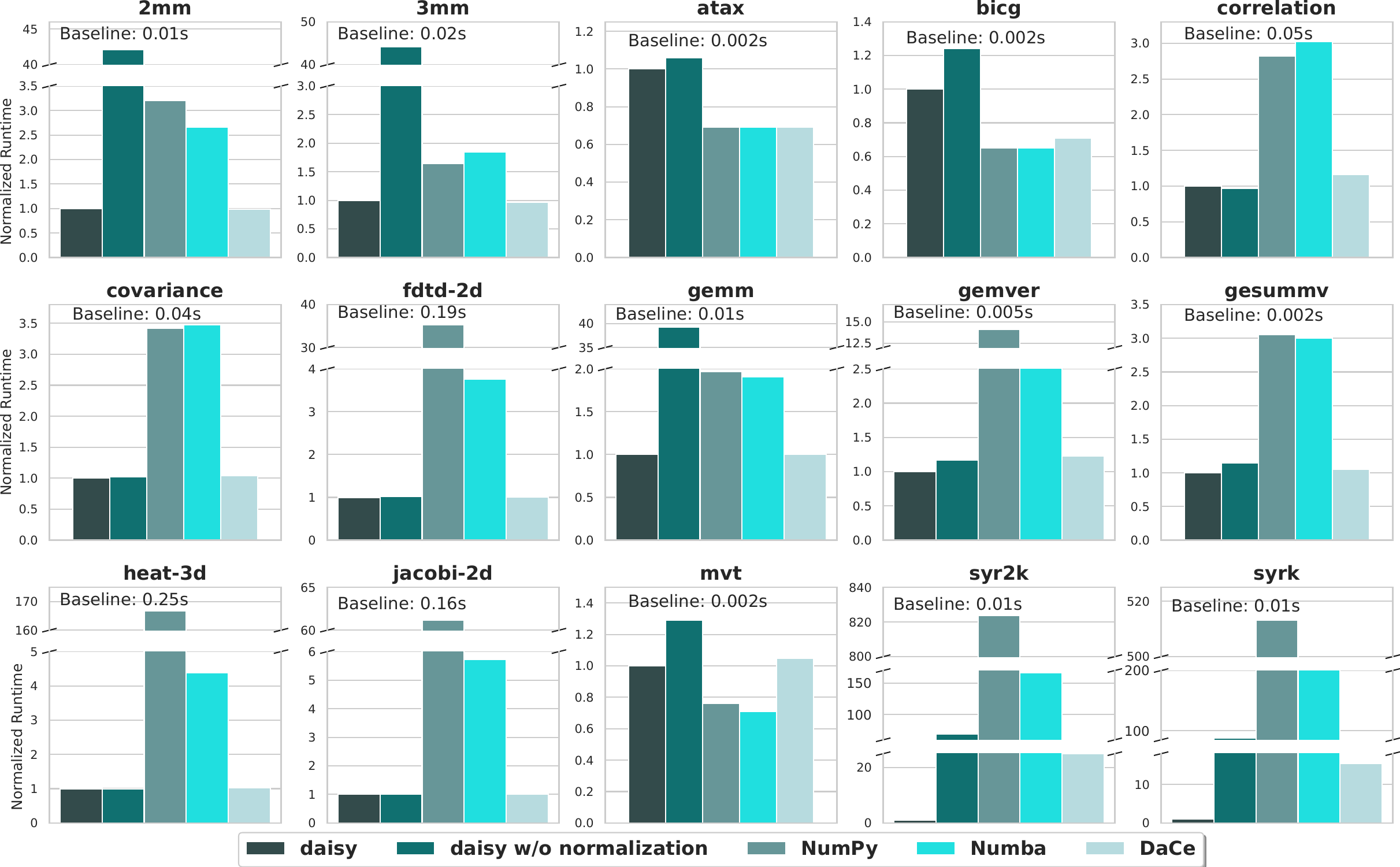}
    \caption{Comparison of our model with NumPy-based frameworks implementing custom operators and optimizations for different applications. The runtime is expressed relative to the runtime of \toolname. Hence, a lower value is better.}
    \label{fig:NPBench}
\end{figure*}
\paragraph{Baselines}
We consider three baselines: NumPy 1.25.2~\cite{Harris:2020}, Numba 0.58.0~\cite{Lam:2015}, and DaCe 0.14.2~\cite{BenNun:2019}.
All frameworks use custom operators to call optimized BLAS libraries for specific operations.
Besides the operators, Numba and DaCe support additional optimizations.
In detail, Numba is a just-in-time compiler that can automatically parallelize and vectorize loops if detected accordingly.
DaCe generates SDFGs from a Python frontend, which can be optimized with various transformations - including, but not limited to the automatic parallelization and vectorization of loops.

\paragraph{Results}
Figure~\ref{fig:NPBench} shows the runtime of the benchmarks for our approach and the improvements over the baseline frameworks.
It also shows the results for \toolname without prior normalization.
Several benchmarks consist of BLAS kernels such as \textit{gemm} and \textit{gemv}, for which the different frameworks provide custom operators.
%We summarized these benchmarks in the first group of the table.
Our model lifts BLAS-3 kernels to matching library calls and optimizes the remaining loop nests using optimizations found with the evolutionary search.
The lifting of BLAS-3 kernels fails without normalization on several benchmarks, e.g., \textit{2mm}, \textit{3mm} and \textit{gemm}.
When applying normalization, our model mostly matches the performance of DaCe and outperforms NumPy and Numba.
For the \textit{syrk} and \textit{syr2k} loop nests, our model outperforms all frameworks because the baseline frameworks do not provide custom operators here.
Compared to unoptimized C code, the native Python code means a significant loss in performance in these cases.
Furthermore, there are benchmarks for which auto-parallelization is necessary to achieve speedups, e.g., \textit{heat-3d} and \textit{jacobi-2d}.
However, the structure of the loops, as implemented by the developer, does not comprise much potential for further optimization.
Note that the correlation and covariance benchmarks do not show the problems of Section~\ref{subsec:Normalized_Scheduling} due to a different structure of the SDFGs from the Python frontend.
The performance of our model is thus close to the performance of DaCe, outperforming DaCe in cases where no custom operators are available.

\textbf{In summary, a priori normalization enables the application of an auto-scheduler derived from specific variants of C loop nests to loop nests translated from Python programs.}

\section{CLOUDSC: Case Study of Normalization and Optimization}
\label{sec:CLOUDSC}

\begin{figure*}
    \centering
    \begin{subfigure}{0.47\textwidth}
        \includegraphics[width=\linewidth]{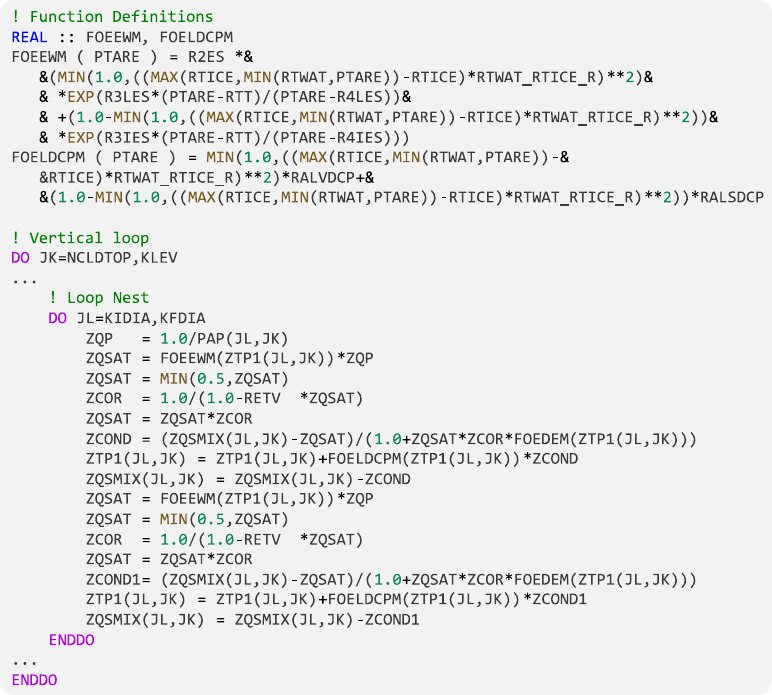}
        \caption{A part of the simulation of the erosion of clouds. Functions definitions are provided, and the outer vertical loop is shown.}
        \label{fig:Cloud_Erosion}
    \end{subfigure}
    \hfill
    \begin{subfigure}{0.46\textwidth}
        \includegraphics[width=\linewidth]{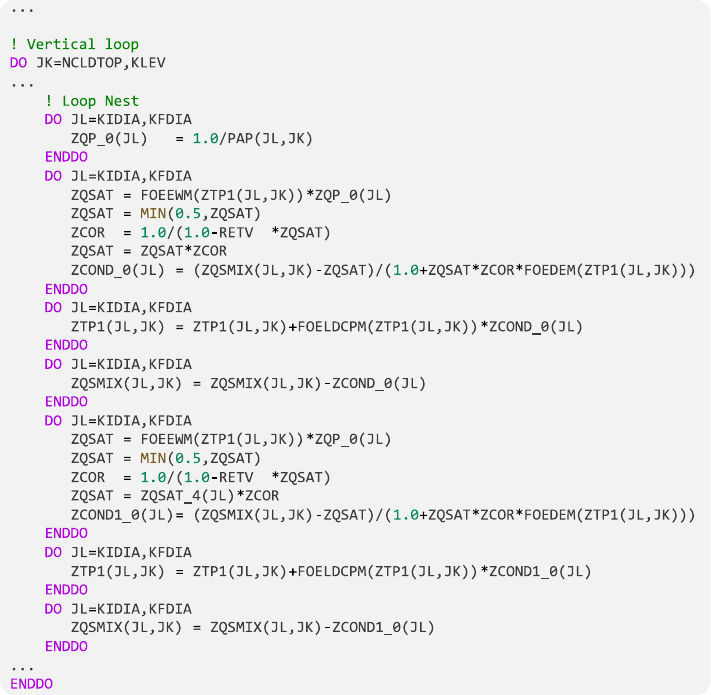}
        \caption{A part of the simulation of the erosion of clouds fissioned into individual computations and fused by one-to-one produce-consumer loop nest relations.}
        \label{fig:Cloud_Erosion_Fission}
        \end{subfigure}
    \caption{A loop nest taken from from the vertical loop of CLOUDSC before and after normalization and fusion.}
    \label{fig:enter-label}
    \vspace{-1em}
\end{figure*}

We now expand our analysis to CLOUDSC, a parametrization scheme for simulating clouds and precipitation.
The model is part of the Integrated Forecasting System (IFS) and is employed in production by ECMWF for weather forecasts and climate analysis.
It is a nonlinear scheme that accounts for approximately 10\% of the IFS forecast model. 
Initially written in Fortran, the code has been the focus of an effort to understand performance portability with versions now existing in C, as well as using OpenACC and CUDA.
We use an SDFG generated by the DaCe framework and apply our normalization pipeline implemented in \textit{daisy}.

An important characteristic of the code is the way it accesses data.
The simulated volume is divided into vertical columns, each computed independently.
When iterating vertically through a column, the physical properties stored in multiple large arrays are updated.
The update during each step of the vertical loop comprises several nested loops, each implementing distinct physical equations.
These innermost loops iterate over a "tiling" parameter --- \texttt{NPROMA}, simultaneously updating multiple independent columns.
Hence, this innermost tiling parameter divides the total number of columns between the outermost loop \texttt{NBLOCKS} and the innermost loops \texttt{NPROMA}, i.e., \texttt{num\_columns=NBLOCKS*NPROMA}.
Since both loops are fully data parallel, users can divide the total problem size into \texttt{NPROMA} and \texttt{NBLOCKS} to find the optimum balance between parallelism and data locality for their hardware system.

\subsection{Discovering Performance Optimizations in the Erosion of Clouds}
\label{subsec:CLOUDSC_Example}

We first analyze a single physical update step inside the vertical loop.
%Inside CLOUDSC, the different physical steps are numbered.
Figure~\ref{fig:Cloud_Erosion} depicts a part of the simulation of cloud erosion.
Since the computation is fully data-oblivious, we can investigate its performance independent of other computations inside the model.

The chosen loop nest is representative of many innermost loops within the vertical loop:
It updates two arrays, \texttt{ZTP1} and \texttt{ZQSMIX}, over the \texttt{NPROMA} dimension with an iterator \texttt{JL} in a fully data-parallel manner.
For this, it computes several intermediate scalars, which can be assumed to be live only for the loop's scope.
By default, CLOUDSC is compiled with loop unrolling and function inlining to maximize instruction-level parallelism.
Therefore, the computations of the sub-routines \texttt{FOEEWM} and \texttt{FOELDCPM} are inlined, and the loop body is unrolled.
Hence, the loop body is significantly larger than the source code suggests, potentially hindering crucial compiler optimizations such as register allocations.

We apply maximal loop fission to divide individual computations into smaller loops.
This reverts the original decision by the developers to group the computations according to the physical equation. 
This enables the application of a typical optimization recipe, which iteratively fuses all one-to-one producer-consumer relations between loop nests.
Hence, intermediate results are computed for the whole dimension of \texttt{JL} and stored in the local arrays \texttt{ZQP\_0} and \texttt{ZCOND\_0}.
Each loop nest only contains scalars, which are used within a short distance of instructions.
The resulting loop nests are shown in Figure~\ref{fig:Cloud_Erosion_Fission}.

\begin{table}[ht]
    \centering
    \begin{tabular}{ l r r}
        & Original & Optimized\\\midrule
        Single Iteration[ms] & 0.040 & 0.006 \\
        \texttt{KLEV} Iterations[ms] & 5.468 & 0.882 \\
        L1 Loads & 2632 & 1281 \\
        L1 Evicts & 963 & 178 \\
        \bottomrule
        \hline
    \end{tabular}
  \caption{The table shows the runtime for a single iteration and \texttt{KLEV} iterations of the loop nests. It further shows the absolute number of loads and evicts on the L1 cache.}
  \label{table:Cloud_Erosion_Table}
\end{table}

\paragraph{Results}
We measure the performance of the two versions for a single iteration and \texttt{KLEV} iterations, corresponding to the size of the vertical loop.
We set \texttt{NPROMA=128}, as this value offers the best results on our hardware.
The experimental setup is analogous to Chapter~\ref{sec:AB_Testing}.
Furthermore, we measure load and evicts to and from the L1 cache to analyze the improvements of our optimization.
Table~\ref{table:Cloud_Erosion_Table} summarizes both runtime and memory behavior, showing that our optimization yields a speedup of $4\times$.
Furthermore, the optimization reduces the pressure on the L1 cache, as there are significantly fewer transfers between L1 and L2.
Hence, the normalization allows us to discover new applications of well-known performance optimizations as it transforms the physical equation into a canonical form.

\subsection{Evaluation}
\label{subsec:CLOUDSC_Model}

\begin{figure}
\centering
    \includegraphics[width=0.6\columnwidth]{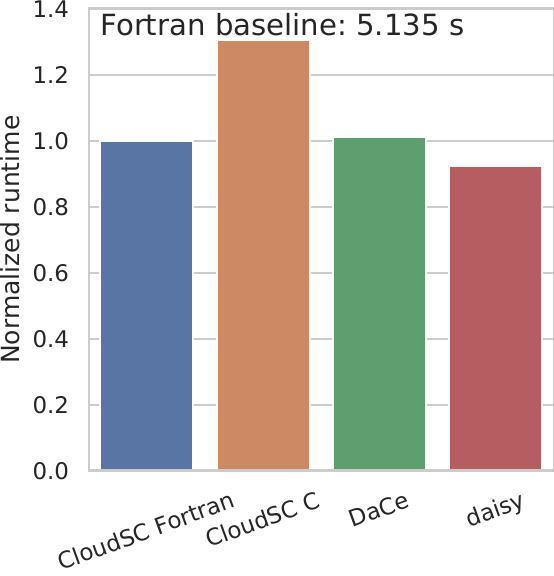}
    \caption{CLOUDSC runtime for sequential execution of the Fortran, C, DaCe, and \toolname versions from left to right. The runtime is normalized by the Fortran version. Hence, a lower value is better.}
    \label{fig:cloudsc_runtime}
\end{figure}
\begin{figure}[ht]
    \centering
    \begin{subfigure}{\linewidth}
    \centering
        \includegraphics[width=\columnwidth]{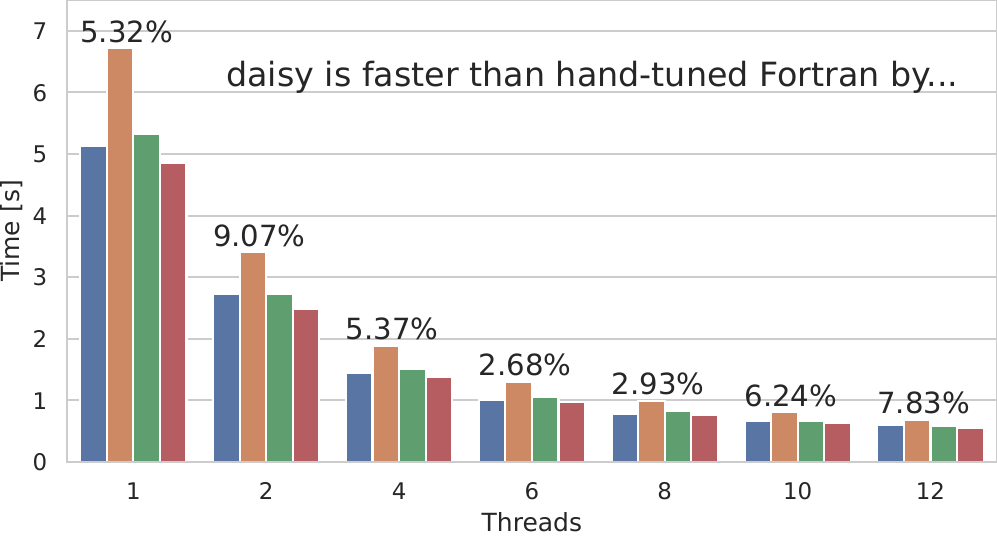}
        \caption{Strong scaling behavior of CLOUDSC for the Fortran, C, DaCe, and daisy versions from left to right and grouped by the number of threads.}
        \label{fig:CloudSC_strong_scaling}
    \end{subfigure}
    \begin{subfigure}{\linewidth}
    \centering
        \includegraphics[width=\columnwidth]{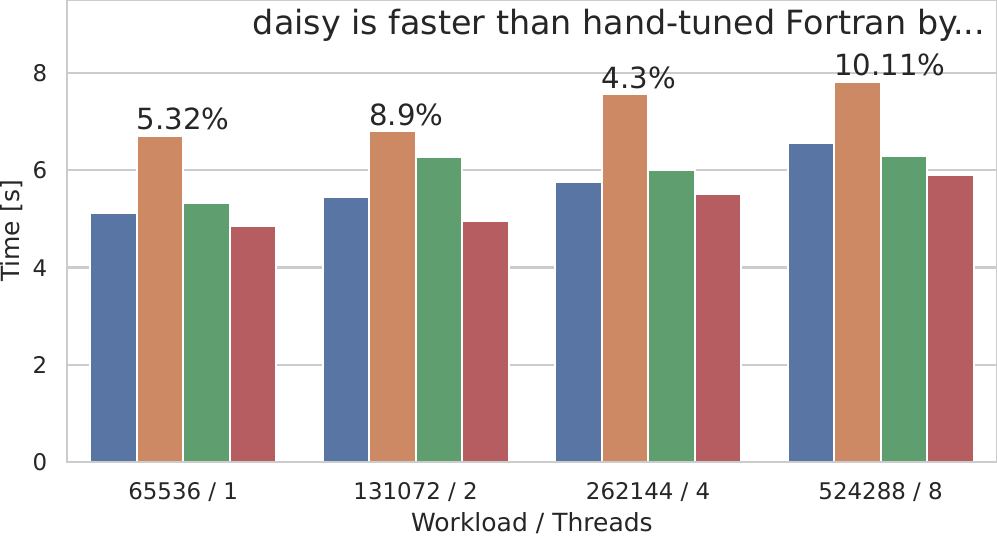}
        \caption{Weak scaling behavior of CLOUDSC for the Fortran, C, DaCe, and \toolname versions.}
        \label{fig:CloudSC_weak_scaling}
    \end{subfigure}
    \caption{Experimental results for the CLOUDSC case study.}
    \label{fig:enter-label}
\end{figure}

We now apply the same pipeline to the full model and compare the runtime of \toolname against the Fortran, C, and DaCe versions.
Figure~\ref{fig:cloudsc_runtime} shows the measurements for sequential execution.
We keep $NPROMA=128$ and set $NBLOCKS=512$.
The optimizations applied by \toolname yield a speedup of 1.08x compared to the second-best version, which is the highly-tuned Fortran code.
As shown previously, optimizing the loop nest in the erosion of clouds saves approximately four milliseconds per execution of the vertical loop, accumulating to about 200 milliseconds for the full model.
Hence, this optimization primarily explains the performance improvements.
Furthermore, we compare the different versions' strong and weak scaling behavior in Figure \ref{fig:CloudSC_strong_scaling} and Figure~\ref{fig:CloudSC_weak_scaling}.
With an optimization of the application's critical path, the performance improvements also translate to the parallel execution.
We also measure the FLOP/s for Fortran and \toolname and compare it to the peak FLOP/s of the machine.
The peak FLOP/s of the machine is measured with 52522.83 MFLOP/s based on a benchmark optimized for Fused-multiply-add (FMA) and AVX instructions.
The Fortran version yields 13634.03 MFLOP/s, and the \toolname version yields 14792.81 MFLOP/s, which is 25.96 \% and 28.16\% of peak performance, respectively.
This underscores that CLOUDSC is already a highly tuned application, so finding optimization opportunities is difficult.
Hence, \toolname improved the performance of a critical loop nest in a highly-tuned application.

\section{Discussion}
\label{sec:Discussion}

The following sections discuss the implications and limitations of our work.

\paragraph{Normalization Criteria}
The idea of our normalization is the simplification of memory accesses.
This is motivated by the observation that optimization recipes usually apply to a particular memory access pattern, but large applications group computations in loop nests according to formulas.
We derived the normalization criteria from two common ways to change the memory accesses of a loop nest: by permutation and composition.
Our findings open a research avenue in exploring normalization criteria and understanding their impact on optimization pipeline performance.
% It is an interesting question whether other normalization criteria, e.g., enforcing untiled loops, may be helpful to apply optimization recipes in complex applications.

\paragraph{Complexity of Auto-scheduling}
Auto-scheduling is typically formulated as a search over a humongous space of possible combinations of transformations~\cite{Adams:2019,Baghdadi:2021,Steiner:2021}.
This unconstrained formulation requires careful design and training of performance models and search methods.
A priori loop nest normalization separates the search and input space and maps semantically equivalent loop nests to a single problem instance.
This reduces the necessary model complexity without a loss of generality of the optimization for the unconstrained search space.

\section{Related Work}
\label{sec:Related_Work}

Loop transformations and normalization have been studied for decades.
In particular, \citet{Chelini:2020} evaluate stride minimization as an optimization criterion for loop scheduling, and \citet{Callahan:1992} discusses loop distribution as a technique to simplify the analysis of parallel loops.
This paper, in turn, demonstrates the application of such techniques for the design of robust auto-schedulers.
The following briefly overviews the related work on auto-scheduling and loop normalization.

\paragraph{Optimization Criteria}
Auto-scheduling approaches define loop scheduling as an optimization problem over different types of objective functions.
In the domain of polyhedral optimization, the minimization of the \textit{maximal dependence distance}~\cite{Lim:1999, Bondhugula:2008, Acharya:2015} is a popular objective function, which is also implemented in \textit{Pluto}~\cite{Bondhugula:2008b} and Polly~\cite{Grosser:2011}.
While these approaches usually guarantee global optima for the solutions of the scheduling problems, the modeled schedule space and objective function may fail to capture the complex features of modern hardware~\cite{Baghdadi:2019}.
Recent works~\cite{Kong:2019, Chelini:2020, Miller:2021} therefore combine multiple specialized criteria such as \textit{Stride Optimization} and \textit{Dependence Guided Fusion} into a multi-objective optimization.
Another research direction learns the objective function from data using deep learning~\cite{Chen:2018,Adams:2019,Baghdadi:2021,Steiner:2021,Singh:2022}.
All these approaches seek to optimize more complex objective functions in larger schedule spaces at the cost of a global optimum.
This paper proposes normalization as a pre-processing step based on criteria that resemble manually derived objective functions.
Normalization is primarily applicable to models based on local optimization to reduce the variation of states.

\paragraph{Idiom Detection}
Idiom detection seeks to detect and replace specific idioms with optimized implementations.
Since the detection is prone to variations, loop normalization techniques have been analyzed in this context early on.
\citet{Pinter:1994} define a loop normal-form for auto-parallelization based on idiom detection, which requires that each loop-carried dependency is between two loop iterations.
The authors propose \textit{loop unrolling} to ensure this property.
\citet{Callahan:1992} analyzes complex bounded recurrences to recognize parallelism in loops.
The author discusses loop distribution as a technique to simplify this analysis.
Several approaches seek to detect idioms on LLVM IR~\cite{Chelini:2019, Ginsbach:2020, DeCarvalho:2021}.
\textit{Declarative Loop Tactics}~\cite{Chelini:2019} uses Polly to compute the Scops of the code and detects idioms based on the polyhedral representation.
This approach enforces a description of memory accesses by affine function, which is more of a constraint than a normalization.
\textit{LiLAC}~\cite{Ginsbach:2020} and \textit{KernelFaRer}~\cite{DeCarvalho:2021} require the standard compiler optimizations to be performed before detection.
In contrast, we propose performing data-centric normalizations beyond the standard optimizations of the compiler, which may significantly change the memory access scheme.

\section{Conclusion}
\label{sec:Conclusion}

In this paper, we present a priori loop nest normalization, simplifying the auto-scheduling of loop nests in complex applications.
The approach maps loop nests with different memory access patterns to the same canonical form, significantly reducing the variety of loop nests to be optimized.

We demonstrate the approach in different case studies, highlighting the improved robustness and higher applicability of optimizations.
The approach outperforms state-of-the-art compilers, auto-schedulers, and performance-oriented frameworks in C, Python, and Fortran by significant factors.
In particular, we apply the approach to a highly-tuned scientific simulation, where it identifies additional optimizations resulting in a 10\% speedup.

The approach enables the application of state-of-the-art auto-schedulers to large scientific code bases, bringing mathematical formulas into a form more amenable to optimization.

\begin{acks}
This project has received funding from the European Research Council (ERC) under the European Union’s Horizon 2020 program (grant agreement PSAP, No. 101002047), from the European High-Performance Computing Joint Undertaking (JU) under grant agreement EUPilot No 101034126 and by the ETH Future Computing Laboratory (EFCL), financed by a donation from Huawei Technologies. L.T. wishes to thank D.K. for many valuable discussions.
\end{acks}

\bibliographystyle{ACM-Reference-Format}
\bibliography{literature}

%\appendix

\end{document}